\documentclass[aps,prx,reprint,groupedaddress]{revtex4-2}
\usepackage{dcolumn}
\usepackage{bm}
\pdfoutput=1
\usepackage[utf8]{inputenc}
\usepackage[T1]{fontenc}
\usepackage{lmodern}
\usepackage{siunitx}
\usepackage{graphicx}
\usepackage{longtable}
\usepackage{tabularx}
\usepackage{braket}
\usepackage[overload]{empheq}
\DeclareSIUnit\gal{Gal}
\DeclareSIUnit\gauss{G}
\DeclareSIUnit\eotvos{E}
\DeclareSIUnit\shot{shot}
\DeclareSIUnit\atoms{atoms}
\DeclareUnicodeCharacter{2212}{-}
\usepackage[english]{babel}
\usepackage{amsmath}
\usepackage{hyperref}

\usepackage[dvipsnames]{xcolor}
\usepackage{tikz}

\begin{document}

\title{A compact differential gravimeter at the quantum projection noise limit}

\author{Camille Janvier}
\author{Vincent Ménoret}
\affiliation{iXblue quantum sensors, Talence, 33400, France}
\author{Sébastien Merlet}
\author{Arnaud Landragin}
\author{Franck Pereira dos Santos}
\affiliation{LNE-SYRTE, Observatoire de Paris, Université PSL, CNRS, Sorbonne Université, Paris, F–75014, France}
\author{Bruno Desruelle}
\affiliation{iXblue quantum sensors, Talence, 33400, France}

\begin{abstract}
Atom interferometry offers new perspectives for geophysics and inertial sensing. We present the industrial prototype of a new type of quantum-based instrument: a compact, transportable, differential quantum gravimeter capable of measuring simultaneously the absolute values of both gravitational acceleration, $g$, and its vertical gradient, $\Gamma_{zz}$. While the sensitivity to $g$ is competitive with the best industrial gravimeters, the sensitivity on $\Gamma_{zz}$ reaches the limit set by quantum projection noise---leading to an unprecedented long-term stability of \SI{0.1}{\eotvos} (\SI{1}{\eotvos}$ = $\SI[tight-spacing = true]{1e-9}{\per\second\squared}). This unique, dual-purpose instrument, paves the way for new applications in geophysics, civil engineering, and gravity-aided navigation, where accurate mapping of the gravitational field plays an important role.
\end{abstract}
\maketitle

\section{Introduction}
Absolute gravimeters are prominent tools in geophysics \cite{vancampGeophysicsTerrestrialTimeVariable2017}, they provide an integrative estimation of the surrounding mass density by measuring the gravitational acceleration $g$ to a very high precision. Transportable \cite{niebauerNewGenerationAbsolute1995,menoretGravityMeasurements1092018}, field-deployable \cite{makinenUseA10020Gravimeter2010,cookeFirstEvaluationAbsolute2021} commercial instruments, based on either "classical" or on "quantum technologies" are commonly used for gravity surveys by end-users in geophysics. However, these instruments remain largely limited by vibration noise from the environment. Gravity gradiometers circumvent this issue by measuring a differential quantity that is fundamentally insensitive to vibrations. This has made them powerful tools for space-borne \cite{rummelGOCEGravitationalGradiometry2011}, and air-borne surveys---in the form of commercial grade relative gravity gradiometers for the latter \cite{dransfieldAirborneGravityGradiometry2007,moodySuperconductingGravityGradiometer2011}. However, these instruments intrinsically suffer from drifts and require calibration on a regular basis. \\
\begin{figure}[!h]
\includegraphics[width=0.9\columnwidth]{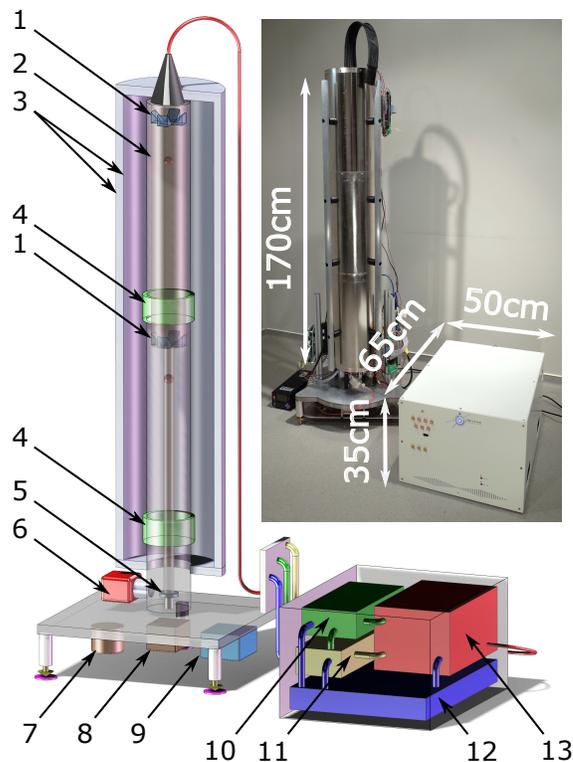}
\caption{Schematics and picture of the instrument: On the left the sensor head, on the right the electronic and laser system. $1)$ Top and bottom trapping pyramidal retro-reflectors. $2)$ Vacuum chamber. $3)$ Inner, outer magnetic shields and coils. $4)$ Top and bottom detection areas. $5)$ Retro-reflecting mirror on a tip-tilt piezo mount. $6)$ Ion pump. $7)$ Barometer. $8)$ Accelerometer. $9)$ Tiltmeter. $10)$ Embedded control computer. $11)$ RF synthesizer. $12)$ Power supply. $13)$ Laser optics. For a more thorough description of each sub-systems see Methods. \label{System}}
\end{figure}
Atom interferometers (AIs) \cite{geigerHighaccuracyInertialMeasurements2020a} offer new solutions to this issue. In the past 20 years, AIs have evolved from large, complex laboratory research experiments \cite{petersHighprecisionGravityMeasurements2001,fixlerAtomInterferometerMeasurement2007,rosiPrecisionMeasurementNewtonian2014,maoDualmagnetoopticaltrapAtomGravity2021},  to compact instruments \cite{menoretGravityMeasurements1092018} that can be used outside the lab \cite{wuGravitySurveysUsing2019a,cookeFirstEvaluationAbsolute2021}, or on moving platforms \cite{barrettDualMatterwaveInertial2016,bidelAbsoluteAirborneGravimetry2020}. Not only have these quantum sensors demonstrated better performance than their classical counterparts \cite{gillotStabilityComparisonTwo2014,freierMobileQuantumGravity2016}, but they offer the possibility to perform simultaneous absolute measurements of the acceleration $g$ and gradient $\Gamma_{zz}$ due to gravity \cite{caldaniSimultaneousAccurateDetermination2019}. Combined, these two quantities provide an improved picture of the surrounding mass distribution \cite{pajotNoiseReductionJoint2008,yeGeneralizedModelMoho2016}. Here, we report on the results obtained with such an instrument that combines state-of-the-art performances with a compact laser system and physical package allowing for its transport and quick installation. Results of its operation as a stationary device and present a proof-of-principle experiment for mass weighing are presented.

\section{Hardware}
\begin{figure*}[t]
\includegraphics[width=1.9\columnwidth]{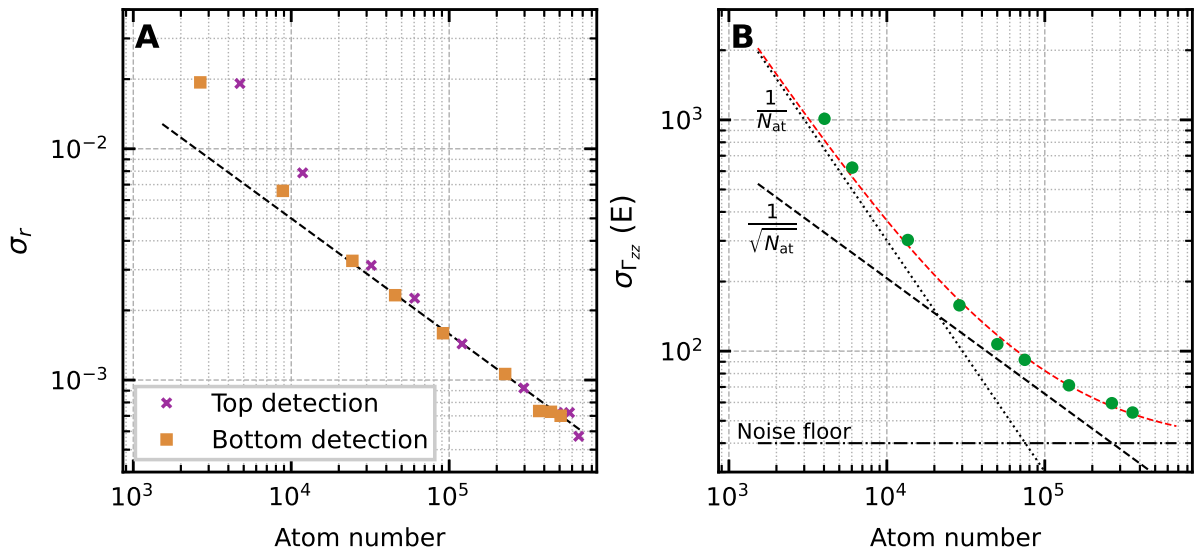}
\caption{Evolution of sensitivity with atom number: Detection (\textbf{A}) and gradient (\textbf{B}) noise as a function of atom number. The QPN model is shown as dashed black line on both panels. On panel \textbf{B} it uses contrasts and parameters from the interferometers to estimate the effect of QPN on $\sigma_{\Gamma_{zz}}$. The full noise model (red dashed line) assumes a level of $1/N_{\rm at}$ noise of $7\times 10^6 \: \rm E\,atom/\sqrt{\tau}$ (dotted line) and a noise floor of \SI{40}{\eotvos} (dash-dotted line). \label{figQPN}  }
\end{figure*}
Benefiting from the experience acquired with the development of Absolute Quantum Gravimeters (AQG), we have developed a transportable compact Differential Quantum Gravimeter (DQG). It uses two vertically-stacked AIs that measure gravity at two different heights while sharing the same interrogation laser. In a similar architecture to the AQG, the instrument is composed of two subsystems: a sensor head which contains the vacuum chamber where the measurement takes place, and an electronics and laser module that generates all the optical and electrical signals necessary for the control of the instrument, see Fig.~\ref{System}. The laser system is based on frequency-doubled telecom fiber lasers and the same proven architecture that was presented in Ref.\cite{menoretGravityMeasurements1092018}, but improved to offer better robustness, a threefold reduction of both volume ---down to \SI{0.1}{\cubic\meter}--- and weight --- down to \SI{33}{\kilogram}. The sensor head is organized around the vacuum chamber and a single laser beam that is used for simultaneous trapping, cooling, and the state manipulation of the two rubidium (Rb) atom clouds (see Methods). It is \SI{175}{\centi\meter} high and weighs \SI{66}{\kilogram}.\\
\begin{figure*}[!ht]
\includegraphics[width=2\columnwidth]{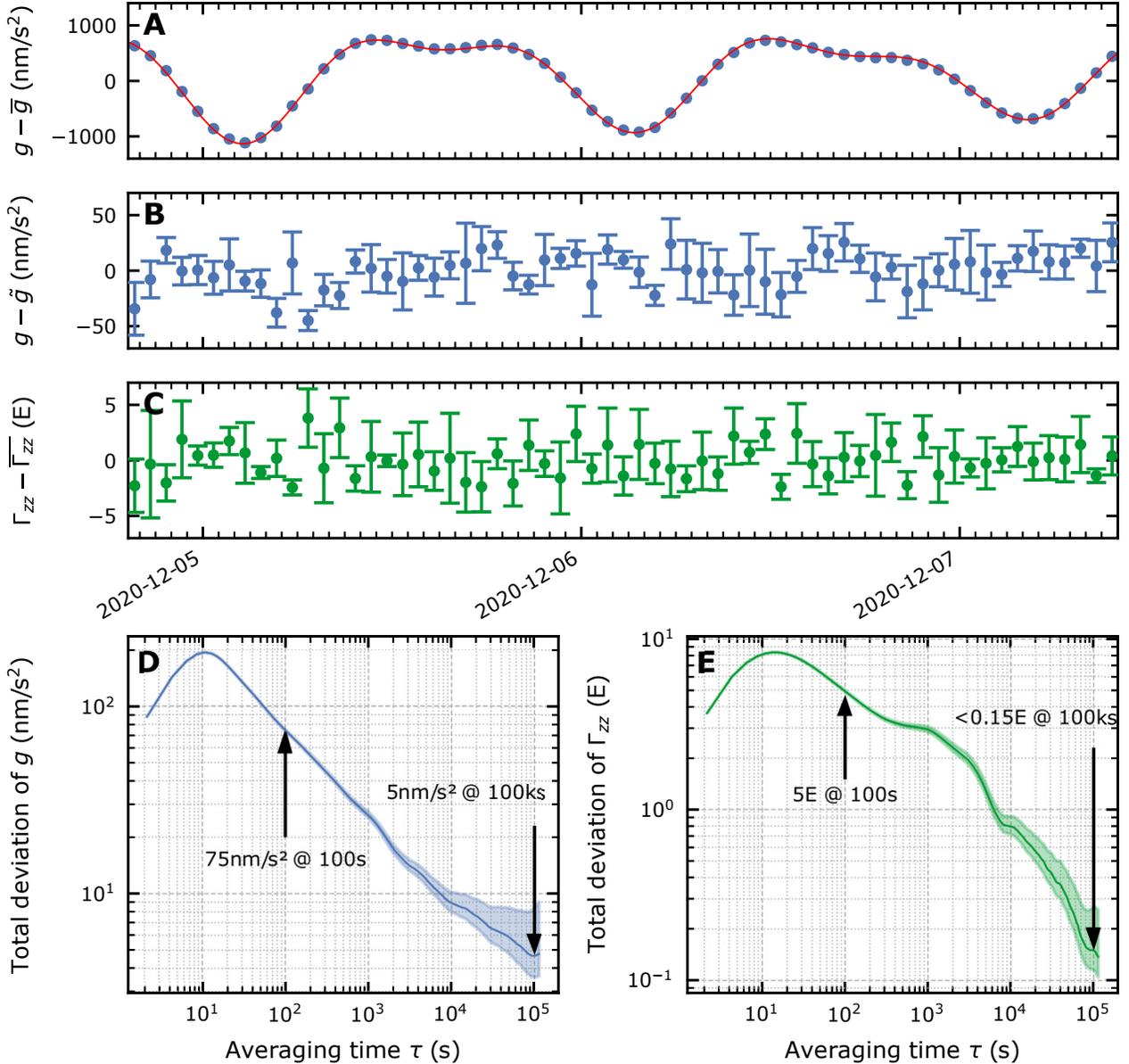}
\caption{Results of a \SI{63}{\hour} dual measurement. (\textbf{A}) Measured $g$ value averaged over 1 hour superimposed with the local tide model (in red). 
(\textbf{B}) Residuals on $g$ averaged over 1 hour after correction for tilt, atmospheric pressure and tides $\tilde{g}$. (\textbf{C}) Measured $\Gamma_{zz}$ value averaged over 1 hour. Average values for $g$ and $\Gamma_{zz}$ over the whole dataset are \SI{9805615664}{\nano\meter\per\second\squared} and \SI{2965.9}{\eotvos}. (\textbf{D}) and (\textbf{E}) total deviation of the residuals of $g$ and $\Gamma_{zz}$ respectively.\label{figTrack}}
\end{figure*}
A measurement cycle lasts \SI{1.08}{\second} and is conducted as follows: two $^{87}$Rb atom clouds are trapped simultaneously using two magneto-optical traps (MOTs) in two pyramidal retro-reflectors separated by \SI{62.5}{\centi\meter} \cite{desruelleColdAtomGravity2014}. Each MOT is loaded for \SI{630}{\milli\second}, after which the magnetic coils are switched off and the atoms further cooled down to \SI{1.5}{\micro\kelvin} before being released. The magnetically sensitive atoms are removed from the cloud using magnetic sub-state selection. After this preparation, a Mach-Zehnder AI of duration $2T=\SI{240}{\milli\second}$ is performed simultaneously on both clouds. The single laser beam produces the stimulated Raman transitions used as the atomic beam splitters \cite{bordeAtomicInterferometryInternal1989,kasevichAtomicInterferometryUsing1991,petersHighprecisionGravityMeasurements2001} and is retro-reflected on a common mirror placed on a piezo tip-tilt mount used to compensate for the Coriolis effect. Both are placed under ultra-high vacuum in order to reduce biases from differential wave-front aberrations between the direct and the retro-reflected Raman beams \cite{louchet-chauvetInfluenceTransverseMotion2011a}. The phase $\phi$ of each interferometer includes the gravitational acceleration integrated over the respective trajectory of each atom cloud. It is extracted by evaluating the transition probability $P\propto \frac{1}{2} C\cos{\phi}$, where $C$ is the interferometer contrast, from the ratio $r$ of atoms detected in each internal state at the output of either interferometer. In order to be maximally sensitive to variations of this phase the interferometers are interrogated at mid-fringe and maintained in this configuration with a dual feedback loop on both the Raman frequency, and on a frequency jump on the Raman detuning for the second pulse of the interferometer, following Ref. \cite{caldaniSimultaneousAccurateDetermination2019}. The feedback on those quantities is then expressed in terms of gravity acceleration $g$, and its vertical gravity gradient $\Gamma_{zz}$ using precisely known parameters of the interferometers (see Methods).

\section{QPN-limited gradient measurement}
One of the main advantages of the differential gravimeter configuration is that a considerable amount of the noise to which gravimeters are sensitive to, is suppressed in the differential signal thanks to efficient common-mode rejection. As a consequence, although the measurement of $g$ is still sensitive to mirror vibrations, laser phase noise, and intensity fluctuations of the laser beam, the measurement of the gradient can become solely limited by the detection noise on the population ratios. This detection noise can be characterized by its Allan deviation $\sigma_r$, and is limited by quantum projection noise (QPN), as we show below. Note that only very few high performance AI instruments have demonstrated operation at the QPN limit \cite{gauguetCharacterizationLimitsColdatom2009,sorrentinoSensitivityLimitsRaman2014}.\\
Detection noise is commonly decomposed in three kinds of contribution to $\sigma_r$, with different scalings with respect to the atom number $N_{\rm at}$: $1/N_{\rm at}$, $1/\sqrt{N_{\rm at}}$, or independent of $N_{\rm at}$ \cite{itanoQuantumProjectionNoise1993,santarelliQuantumProjectionNoise1999}. The first contribution is dominated by technical noise such as shot noise in the electronics or stray light fluctuations. The second is the quantum projection noise which is a manifestation of the probabilistic nature of the measurement of a quantum superposition. The third arises for instance from optical noise due to laser frequency and intensity fluctuations. In the following, we first assess detection noise independently of any interferometer by preparing a state superposition using a $\pi/2$ microwave pulse. The result of this measurement is shown Fig.~\ref{figQPN} \textbf{A}. We observe that the detection noise decreases for both the top and bottom clouds with respect to the detected atom number. While technical noise dominates at low atom number with a higher contribution for the top cloud, this contribution becomes negligible above $5 \times 10^4$ atoms for both clouds. Above this number, the two datasets follow the same $1/\sqrt{N_{\rm at}}$ scaling---confirming that the detection noise is indeed limited by QPN. No trace of a noise floor is observed in these data. \\
To confirm that the gravity gradient measurement is limited by detection noise and therefore by QPN, we measure the differential gravity noise $\sigma_{\Gamma_{zz}}$ as a function of atom number. This quantity is estimated from the difference between detection ratios obtained from the top and bottom interferometers while both are in phase and at mid-fringe. Respective contrasts were measured to be $C_{\rm bottom}=0.53$ and $C_{\rm top}=0.42$. The result is shown Fig.~\ref{figQPN} \textbf{B}. As for the detection noise, the differential gravity noise is limited by technical noise at low atom number. For atom numbers ranging from $2.5 \times 10^4$ to $2.5 \times 10^5$, $\sigma_{\Gamma_{zz}}$ is dominated by $1/\sqrt{N_{\rm at}}$ noise which is well modeled by QPN accounting for the respective contrasts of the interferometers. In this regard, both datasets are in agreement and decreasing as the square-root of the atom number for approximately one decade---confirming that the gradient measurement is indeed QPN-limited in the operating range below $2.5 \times 10^5$ atoms. Above this number, a fit to these data provides an estimated noise floor of \SI{40}{\eotvos}. This is on the same level as the best sensitivity reported for an AI gravity gradiometer \cite{asenbaumPhaseShiftAtom2017}. We attribute this noise floor to frequency noise on the Raman lasers \cite{legouetInfluenceLasersPropagation2007}.  

\section{Stability of the dual measurement}
Having characterized the short-term differential sensitivity, we now focus on long-term measurements. We present on Fig.~\ref{figTrack} a 63 h-long differential gravity measurement obtained in laboratory conditions. We calculate gravity residuals $g-\tilde{g}$ by correcting the raw gravity signal for tilt \cite{niebauerOfflevelCorrectionsGravity2016}, atmospheric pressure and tidal fluctuations using a bespoke model for our measurement site \cite{wziontekStatusInternationalGravity2021}. The residuals reveal no significant drift which is confirmed by the total deviation that continuously decreases with averaging times down to \SI{5}{\nano\meter\per\second\squared} at \SI{10000}{\second}. From the slope of the total deviation we estimate the sensitivity of the gravity measurement at \SI{750}{\nano\meter\per\second\squared\per\sqrt{\tau}}. Unlike the gravity gradient measurement, this sensitivity is typically limited in our urban environment by acoustic and seismic noise due to anthropic activities. We mitigate these effects using an active vibration compensation system, which uses the signal of a classical accelerometer to act directly on the laser phase during the interferometer \cite{lautierHybridizingMatterwaveClassical2014}, as well as rubber pads placed under the apparatus to reduce high frequency noise.\\
The total deviation of the gravity gradient measurement reaches \SI{5}{\eotvos} after \SI{100}{\second} of integration which is compatible to the sensitivity measurements presented in Fig.~\ref{figQPN}. The noise averages down with a hump around \SI{1000}{\second}, suggesting hourly fluctuations, down to \SI{0.15}{\eotvos} at \SI{110000}{\second}. To our knowledge this is presently the best reported stability for a gravity gradiometer \cite{sorrentinoSensitivityLimitsRaman2014}. In terms of differential gravity measurement this represents a difference of less than \SI{100}{\pico\meter\per\second\squared} between the two interferometers. In terms of detectability, it corresponds to the gravity gradient anomaly generated by a \SI{1}{\liter} cubical void in the ground \SI{37}{\centi\meter} directly under the instrument (see Methods).

\begin{figure}[t]
\includegraphics[width=0.9\columnwidth]{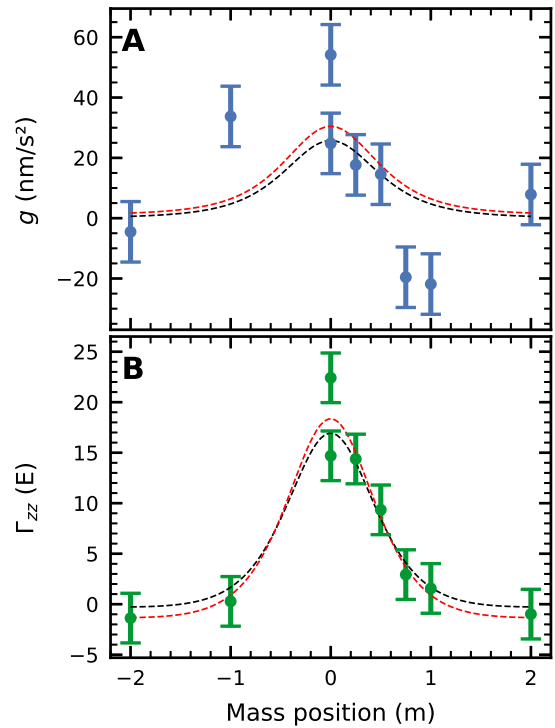}
\caption{Signals acquired by the DQG as a 147\,kg lead mass is moved below the instrument with a 1 hour integration duration per point. While the signal on $g$ (\textbf{A}) is not resolved, the signal on $\Gamma_{zz}$ (\textbf{B}) clearly reveals the effect of the mass. Error bars were estimated from the total deviation of a long measurement acquired the night after the experiment assuming a stationary noise. The model (black dashed lines) uses the physical parameters of the mass as well as the interferometer geometries to provide an accurate evaluation of the theoretical signal, the only adjustable parameters are the offsets of the data (see Methods). The fit (red dashed lines) adjusts for density, gradient and position offset on the gradient data. \label{figMasses}}
\end{figure}

\section{Mass weighing experiment}
In order to further illustrate the potential for gravity surveys we perform a mass detection and weighing experiment. Compared to experiments that aim at accurately measuring the Newtonian constant of gravitation $G$ \cite{fixlerAtomInterferometerMeasurement2007,rosiPrecisionMeasurementNewtonian2014,maoDualmagnetoopticaltrapAtomGravity2021}, we chose application-relevant measurement parameters in terms of acquisition duration, size of the mass and its position with respect to the instrument. The sensor head was lifted \SI{16}{\centi\meter} above the ground on a table and a \SI{147}{\kilogram} lead mass was progressively moved below the table and then further away with an integration time of 1 hour per position. The result is shown figure \ref{figMasses}, the effect of the mass on the gradient is clearly resolved with a signal-to-noise ratio of $\sim 10$, while the sensitivity on $g$ is insufficient to resolve the anomaly with such a short averaging time. The experimental data is consistent with theoretical calculations, that take into account the geometry of the mass and the configuration of the interferometers (see Methods). Furthermore, in a more applicative approach, we can estimate the mass or density of the object from the data by fitting the gravity gradient anomaly using a least-squares adjustment of our theoretical model. We assume the size and dimension of the object to be known which is still relevant on an application standpoint. Indeed, in geophysics or civil engineering complementary methods such as ground-penetrating radar can provide this type of information. The fit estimate gives an estimated mass of \SI{168(17)}{\kilo\gram} which is whithin a two-sigma uncertainty of the actual mass.  We note that the gravity signal is here too small to be used, but that for a larger mass it would be useful to better constrain the fit or improve its accuracy.\\

\section{Conclusion and perspectives}
We have presented a new generation of AI gravity sensor that realizes quantum-limited measurements while remaining compact and light enough to be handled by two people. We demonstrated state-of-the-art sensitivity and long-term performance of both gravitational acceleration and vertical gravity gradient in a controlled laboratory environment, as well as its potential for rapid detection and weighing of gravity anomalies such as subterranean masses or voids. This dual-purpose instrument---the first of its kind---is a prototype for a field-deployable device, yet it can already be employed for stationary geophysics measurements. This already constitutes a true leap forward from massive laboratory experiments with limited up-time and fundamental research goals. In the near future, a full accuracy budget will establish the DQG as an absolute instrument for both $g$ and $\Gamma_{zz}$. Furthermore, improvements in terms of ruggedization will lead to an operational instrument for outdoor use with practical applications for geophysics, civil engineering, and gravity-aided navigation. \\

\section{Acknowledgments}
We would like to thank all the technical, scientific and administrative personnel of Muquans for their work and support. We also thank Jackie Johnstone and Brynle Barret for their valuable input on the paper. We acknowledge financial support by the ANR under contract ANR-19-CE47-0003 GRADUS, and DGA under contract 162906044 GRADIOM.\\

%

\newpage
\appendix

\section{Material and methods}
\subsection{Hardware description of the instrument}
\paragraph{Laser and electronics system:}
Two lasers are used to trap and manipulate the atoms in the sensor head. They are generated from frequency-doubled \SI{1560}{\nano\meter} Extended Cavity Diode Lasers (ECDLs) and amplified by two Erbium-Doped Fiber Amplifiers (EDFAs). The two ECDL are phase-locked onto a master laser locked on a rubidium spectral line using a saturated absorption scheme. We use commercial periodically-poled lithium niobate (PPLN) doubling crystals for the frequency doubling. An acousto-optical-modulator (AOM) placed in the laser system generates the laser pulses.
Except for the master laser, all frequencies in the instrument are generated from a \SI{100}{\mega\hertz} oven-controlled crystal oscillator. Intermediate frequencies are synthesized from this reference using a \SI{7}{\giga\hertz}  phase-locked dielectric resonator oscillator (PLDRO) and direct digital synthesizers (DDS). A measurement cycle is managed by an embedded computer in the laser system. This computer directly controls all the subsystems (including DDS, AOM, EDFA power) and gathers all the data from the sensor head (detection signals, tiltmeter, etc.) and performs all the calculations necessary to the dual tracking sequence. The measurement output and monitoring data is streamed by this on-board computer to a user interface computer used for remote control of the instrument and data collection.

\paragraph{Sensor head:}
The sensor head is organized around a free-standing titanium vacuum chamber and is about \SI{175}{\centi\meter} high, \SI{55}{\centi\meter} wide and weighs \SI{66}{\kilo\gram}. In order to perform the differential measurement, two atom clouds are trapped in two different pyramidal retro-reflectors separated by \SI{62.5}{\centi\meter}. Commercial alkali metal dispensers are used to produce a rubidium vapor in the vicinity of each trap. A notable difference with respect to the AQG design \cite{menoretGravityMeasurements1092018} is the placing of the mirror at the bottom of the vacuum chamber instead of the top in order to accommodate the piezo tip-tilt actuator used for Coriolis effect compensation \cite{petersHighprecisionGravityMeasurements2001,freierAtomInterferometryGeodetic2017}. The retro-reflecting mirror that serves as an inertial reference frame for the measurement is fixed on top of this piezo-actuator used to rotate the mirror during the free fall to compensate for Coriolis effect. Both are placed under ultra-high vacuum in order to reduce biases from differential wave-front aberrations between the direct and the retro-reflected Raman beams \cite{louchet-chauvetInfluenceTransverseMotion2011a}. Each atom cloud is detected in a dedicated detection zone consisting of two rows of photodiodes. The vacuum chamber is rigidly bolted to an aluminum tripod and supports two bias coils that generate a uniform magnetic field of \SI{8}{\micro\tesla} along the vertical axis, two layers of magnetic shielding and the laser collimator placed on top of it. Ultra-high vacuum inside the chamber is maintained using non-evaporable getters and an ion pump. All optical signals are amplified and digitized directly on the sensor head. A control card is responsible for the accelerometer signal acquisition and processing, involved in the active vibration compensation \cite{menoretGravityMeasurements1092018}. It also manages and synchronizes the sensor head subsystems and the communication with the laser system. A tiltmeter and a barometer complete the sensor instrumentation.

\subsection{Dual tracking}
\label{Dual tracking}
Quantum gravimeters measure $g$ from the phase of an atomic Mach-Zender interferometer by locking the Raman laser on its central fringe using an adjustable frequency chirp. To be able to track two interferometers at the same time, Caldani \textit{et al.} \cite{caldaniSimultaneousAccurateDetermination2019} proposed to use an adjustment of the laser wave-vector \cite{rouraCircumventingHeisenbergUncertainty2017} to control the phase difference between the two interferometers. Adjusting both the frequency chirp and the wave-vector gives access to $g$ and $\Gamma_{zz}$ simultaneously. It also ensures a maximal common mode rejection \cite{dossantosDifferentialPhaseExtraction2015}, reduced sensitivity to contrast fluctuations, and optimizes the sensitivity of the instrument to phase variations of the interferometers by always measuring on the highest slope of the fringe. More importantly this method eliminates the need for an independent measurement of the distance between the clouds, and is insensitive to its fluctuations. 

In order to compensate for the effects of magnetic field and first order light shifts, the effective wave vector of the Raman transition is reversed between successive measurement cycles---leading to two interleaved and independent dual-tracking feedback loops \cite{weissPrecisionMeasurementHbar1994}. Finally, post-corrections are applied to the raw gravity signal to compensate for tides, atmospheric pressure\cite{wziontekStatusInternationalGravity2021}, and tilts of the instrument \cite{niebauerOfflevelCorrectionsGravity2016}

Stability calculations presented in Fig.3 use the total deviation, with error bars computed according to ref \cite{howeTotalDeviationApproach2000}.

\subsection{Detection noise and gradiometer sensitivity}
\paragraph{Detection protocol:} The population ratio between the outputs of the interferometers is measured as follow: a pulse of molasses light, red-detuned from the $\ket{F=2} \to \ket{F'=3}$ cycling transition in $^{87}$Rb, projects the measurement and stops the atoms in the $\ket{F=2}$ state in front of a first row of photodiodes in each of the two detection area, while the atoms in $\ket{F=1}$ continue to fall. Once the $\ket{F=1}$ atoms reach the second row of photodiodes, a resonant laser pulse is applied and the atomic fluorescence is collected by the photodiodes \cite{mcguirkLownoiseDetectionUltracold2001}. Finally a blow-away pulse (blue-detuned from the cycling transition) is applied to remove all the remaining atoms from the detection area, and a second detection pulse is applied to measure and subtract any background light. The detection ratio is calculated from the fluorescence signal corrected by the detection offsets and crosstalks between detection rows. The same signals are converted to atom numbers \cite{steckRubidium87Line2019}, and a calibration factor that was estimated from the QPN measurement Fig.2A) \cite{santarelliQuantumProjectionNoise1999}.

\paragraph{Atom number:} The atom numbers in the sensitivity experiments was controlled by varying the  time the laser was on during the MOT phase. Increasing this duration results in an increased atom number without changes to other experimental parameters. Crosstalks between two rows of photodiodes were measured for each configuration in order to account for the varying shape of the atom cloud. 

\paragraph{Noise calculations:} The detection noise in FIG2 A) was calculated as the 1-sample Allan deviation of the detection ratio and plotted as a function of the respective atom number. For Fig.2B, the differential phase noise was measured at mid-fringe during an interferometer and was calculated as :
\begin{equation}
\sigma_{\Gamma_{zz}}=\frac{2}{L k_{\rm eff}T^2}\left(\frac{r_{\rm top}}{C_{\rm top}}-\frac{r_{\rm bottom}}{C_{\rm bottom}}\right),
\end{equation}
where $L$ is the distance between atom clouds, $k_{\rm eff}$ the Raman effective wave vector, $T$ the free evolution duration of the interferometer, $r_{i}$ and $C_i$ are the population ratio and contrast of the interferometer $i =$ top, bottom. 

\subsection{Setup and simulation of the mass detection}
The sensor head of the DQG is placed on a small platform in order to be able to pass masses directly under it. The masses are moved under and away from it in steps. The distance between the top of the masses and the mirror is \SI{31.5}{\centi\meter}. The mass is made of twenty \SI{7.3}{\kilogram} and $360 \times 90 \times 20$\,$\textrm{mm}^3$ lead bricks for an easier manipulation. The assembled mass weighs \SI{146.7}{\kilogram} and measures $360 \times 360 \times 100$\,$\textrm{mm}^3$.

Because its small size and its density, the mass creates a gravitational attraction that is not linear along the atomic trajectories or between the two clouds. In order to have an accurate estimation of the effect of this mass on the measurement output of the DQG, we have to take into account these variations. To do so, we used a closed-form formula for prismatic masses \cite{liThreeDimensionalGravity1997} to calculate the gravitational pull of the mass along the unperturbed trajectories (\textit{ie} assuming a constant gravitational acceleration) of the atomic wave-packets, taking into account the different Raman recoil velocities in each arm of the interferometers. This force was then numerically integrated using the perturbative approach described in Ref.\cite{dagostinoPerturbationsLocalGravity2011}. This calculation was performed for each interferometer and each effective wave vector orientation. From this we extrapolated the theoretical anomaly both in $g$ and $\Gamma_{zz}$ as a function of the mass's position with respect to the instrument.

The same calculation was used for the estimation of the anomaly generated by a \SI{1}{\liter} void in the ground. The density of the cube was taken to be \SI{2600}{\kg\per\cubic\meter}, directly under the instrument with its top surface \SI{37}{\centi\meter} under the the sensor head. The anomaly was calculated to be \SI{144}{\milli\eotvos} on the gravity gradient and \SI{-0.24}{\nano\meter\per\second\squared} on gravitational acceleration.

\end{document}